\documentclass[12pt, a4paper]{article}
\usepackage{bm}
\usepackage{graphicx}
\usepackage{amssymb,amsmath,amsbsy,amsgen,amsfonts}
\usepackage{dcolumn}
\usepackage{amsthm}
\usepackage{mathrsfs}
\usepackage{latexsym}
\usepackage{array}
\usepackage{amstext}
\usepackage{color}

\usepackage[super,comma,sort&compress]{natbib}

\usepackage{epstopdf}

\newcommand{\bra}[1]{\left\langle{#1}\right\vert}
\newcommand{\ket}[1]{\left\vert{#1}\right\rangle}
\newcommand{\ketbra}[2]{|#1\rangle \langle#2|}

\newcommand{\be}{\begin{equation}}
\newcommand{\ee}{\end{equation}}
\newcommand{\ba}{\begin{array}}
\newcommand{\ea}{\end{array}}
\newcommand{\bqa}{\begin{eqnarray}}
\newcommand{\eqa}{\end{eqnarray}}

\newcommand{\tr}{\mbox{Tr}}

\usepackage[labelsep=endash]{caption}

\newcommand*{\myfontHel}{\fontfamily{phv}\selectfont}

    \makeatletter
    \renewenvironment{thebibliography}[1]{%
          \list{\@biblabel{\@arabic\c@enumiv}}%
               {\settowidth\labelwidth{\@biblabel{#1}}%
                \leftmargin\labelwidth
                \advance\leftmargin\labelsep
                \@openbib@code
                \usecounter{enumiv}%
                \let\p@enumiv\@empty
                \renewcommand\theenumiv{\@arabic\c@enumiv}}%
          \sloppy
          \clubpenalty4000
          \@clubpenalty \clubpenalty
          \widowpenalty4000%
          \sfcode`\.\@m}
         {\def\@noitemerr
           {\@latex@warning{Empty `thebibliography' environment}}%
          \endlist}
    \makeatother

\makeatletter
\renewcommand\@biblabel[1]{#1.}
\makeatother

    \newenvironment{changemargin}[2]{%
  \begin{list}{}{%
    \setlength{\topsep}{0pt}%
    \setlength{\leftmargin}{#1}%
    \setlength{\rightmargin}{#2}%
    \setlength{\listparindent}{\parindent}%
    \setlength{\itemindent}{\parindent}%
    \setlength{\parsep}{\parskip}%
  }%
  \item[]}{\end{list}}

\usepackage[top=1.2in, bottom=1.5in, left=1.2in, right=1.2in]{geometry}

\begin{document}

\vskip-5cm
\noindent
{\Large {\bf {\myfontHel Experimental demonstration of graph-state quantum secret sharing}}}
\vskip0.1cm
\noindent
\begin{flushleft}B. A. Bell$^1$, D. Markham$^2$$^{*}$, D. A. Herrera-Mart\'i$^3$, A. Marin$^2$, W. J. Wadsworth$^4$, J. G. Rarity$^1$ \& M. S. Tame$^{5,6}$$^{*}$\end{flushleft}
\vskip-0.3cm
\noindent
\begin{flushleft}
{\small \it $^1$Centre for Communications Research, Department of Electrical and Electronic Engineering, University of Bristol, Merchant Venturers Building, Woodland Road, Bristol, BS8 1UB, UK \newline \vskip-0.7cm
\noindent
$^2$CNRS LTCI, D\'epartement Informatique et R\'eseaux, Telecom ParisTech, 23 avenue d'Italie, CS 51327, 75214 Paris CEDEX 13, France \newline \vskip-0.7cm
\noindent
$^3$Racah Institute of Physics, The Hebrew University of Jerusalem, Jerusalem 91904 Israel \newline \vskip-0.7cm
\noindent
$^4$Centre for Photonics and Photonic Materials, Department of Physics, University of Bath, Claverton Down, Bath, BA2 7AY, UK \newline \vskip-0.7cm
\noindent
$^5$School\,of\,Chemistry\,and\,Physics,\,University\,of\,KwaZulu-Natal,\,Durban\,4001,\,South\,Africa \newline \vskip-0.7cm
$^6$National Institute for Theoretical Physics, University of KwaZulu-Natal,Durban 4001, South Africa \newline}
 \end{flushleft}

\vskip-0.8cm
\noindent


{\bf Distributed quantum communication and quantum computing offer many new opportunities for quantum information processing. Here networks based on highly nonlocal quantum resources with complex entanglement structures have been proposed for distributing, sharing and processing quantum information. Graph states in particular have emerged as powerful resources for such tasks using measurement-based techniques. We report an experimental demonstration of graph-state quantum secret sharing, an important primitive for a quantum network. We use an all-optical setup to encode quantum information into photons representing a five-qubit graph state. We are able to reliably encode, distribute and share quantum information between four parties. In our experiment we demonstrate the integration of three distinct secret sharing protocols, which allow for security and protocol parameters not possible with any single protocol alone. Our results show that graph states are a promising approach for sophisticated multi-layered protocols in quantum networks.}

\vskip0.5cm
\vskip0.8cm

{\Large {\bf Introduction}}
\vskip0.2cm

The potential benefits of quantum information processing in a connected world are now well established: while the algorithmic speedups offered by quantum computers~\cite{Ladd2010} and the security provided by quantum key distribution~\cite{Gisin2002} are outstanding improvements over what is classically achievable, in recent years many new protocols have emerged in the setting of quantum networks~\cite{Kimble2008}. Examples of these protocols include quantum coin flipping~\cite{Aharanov2000,Spekkens2002,Colbeck2007,Chailloux2009}, blind quantum computation~\cite{Broadbent2008,Bartz2012} and distributed and secure quantum computation~\cite{Buhrman2003,Ben-Or2006}. In this work we investigate the important networking protocol of quantum secret sharing~\cite{HBB99,CGL99} - which allows one party to distribute a secret (classical or quantum) to a network of parties, such that only authorised sets of parties can access the secret and unauthorised sets obtain no information. Secret sharing has many applications in network-based scenarios, such as auctioning, remote voting, money transfer and multiparty secure computation. The first classical protocols for secret sharing were introduced in 1979 by Shamir~\cite{Sha79} and Blakely~\cite{Blakely79}, with quantum versions later developed using quantum channels for sharing both classical and quantum secrets~\cite{HBB99,CGL99}. Most recently, however, secret sharing protocols have been unified under the framework of graph states~\cite{MS08,Keet10,Hein2005} - quantum resources that can be used to share both classical and quantum secrets. One of the most promising features of graph-state based quantum secret sharing is the natural capacity of the resource states to be integrated into more complex networking protocols. Indeed, graph states are also the basis for universal measurement-based quantum computation~\cite{Raussendorf2001,Raussendorf2003,Kashefi05,Browne07,Briegel2009,Mhalla10}, error correction~\cite{GottesmanThesis,Schlingemann2001,Schlingemann2002,Schlingemann2003,Aliferis2006,Dawson2006,Silva2007} and blind quantum computation~\cite{Broadbent2008,Bartz2012}, making them versatile resources for distributed quantum information processing.

\qquad In this work we report an experimental demonstration of graph-state based sharing of classical and quantum secrets using photons in a linear optics setup. We show how a five-qubit graph state can be used for sharing a classical secret amongst four parties using quantum channels (CQ) - secure against a distrusted channel between the dealer (the party that shares the secret) and the four parties. We also outline and demonstrate three protocols of increasing sophistication that allow the same five-qubit graph state to be used to share a quantum secret with quantum channels (QQ). By combining the classical Shamir-Blakely protocols~\cite{Sha79,Blakely79} with CQ and schemes for sharing quantum secrets recently introduced in refs.~\cite{MS08,Keet10,Marin13} we demonstrate the sharing of a quantum secret over an access structure impossible with QQ alone, which is certified as secure against distrusted channels between the dealer and the other parties (also impossible with QQ alone). We thus demonstrate the practical potential of graph-state quantum secret sharing, as well as the capacity for integrating several cryptographic protocols in this setting. Previous experiments on quantum secret sharing have concentrated on sharing classical secrets~\cite{Tittel01,Schmid06,Schmid05,Chen05,Bogdanski08}, with some work regarding the sharing of quantum secrets amongst three players~\cite{TS02,TS03,Lance2003,Lance2004,Lance2005}. However, our work goes beyond these previous works in two crucial aspects. First, the quantum secret sharing is performed using graph states, which are of great importance for the integration of secret sharing with a wide range of quantum networking protocols via the measurement-based paradigm. Second, we have combined three different cryptographic protocols, one classical and two quantum to enable the sharing of a quantum secret which would not have been possible with any one of the protocols alone, thus demonstrating the ability of graph states to allow the hybridisation of classical and quantum protocols. The results of our experimental demonstration and their analysis show some of the key advantages of using graph states for quantum communication protocols in future quantum networks.

\vskip0.5cm
\vskip0.8cm

{\Large {\bf Results}}
\vskip0.4cm

{\bf Resource characterization}

\vskip0.2cm

\begin{figure}[t]
\centering
\includegraphics[width=15cm]{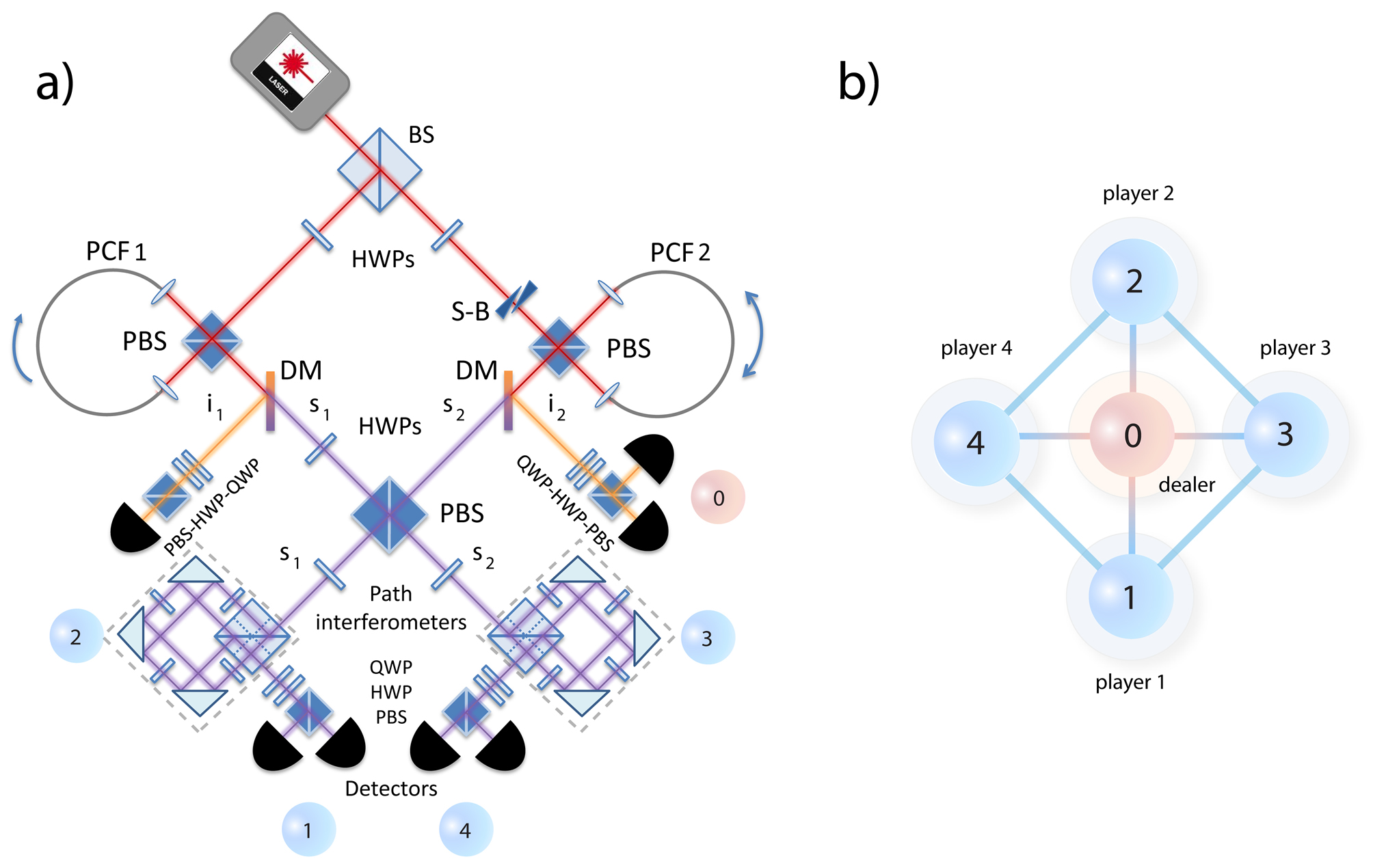}
\vskip1cm
\caption{}
\label{expsetup}
\end{figure}

The setup used to demonstrate graph-state quantum secret sharing is shown in Figure~\ref{expsetup}~a and generates the five-qubit graph state shown in Figure~\ref{expsetup}~b, which acted as a resource state for carrying out the protocols. In the graph state there is initial entanglement between the dealer's qubit (centre qubit) and that of each of the four parties, or players (the outer qubits). The state was generated using the method described in ref.~\cite{Bell13}, where a birefringent photonic crystal fibre (PCF) generates a polarization-entangled pair of photons in the state $\frac{1}{\sqrt{2}}(\ket{H}\ket{H}+\ket{V}\ket{V})$, with $H$ and $V$ referring to horizontal and vertical polarization. The entangled photons are generated at non-degenerate signal and idler wavelengths of $625$nm and $860$nm. A second PCF generates heralded single photons at the signal wavelength (see Methods). Both PCF sources were pumped by the same pulsed laser. The signal photons from the two PCF sources were then overlapped at a polarizing beamsplitter (PBS) to perform a postselected fusion operation~\cite{Bell12}, leaving a three-photon entangled GHZ state $\frac{1}{\sqrt{2}}\left(\ket{HHH}+\ket{VVV}\right)$. This state can be converted by local operations to a linear graph state $\frac{1}{\sqrt{2}}\left(\ket{+0+}+\ket{-1-}\right)$, where the single-qubit computational basis states $\ket{0}$ and $\ket{1}$ are encoded as horizontal and vertical polarizations, and therefore $\ket{\pm}=\frac{1}{\sqrt{2}}\left(\ket{0}\pm\ket{1}\right)$ are encoded as diagonal and anti-diagonal plane polarizations. These $45^\circ$ rotations are applied to the two signal photons emerging from the PBS fusion operation using half-wave plates (HWP).

\qquad Additional qubits are then added to the linear graph state by expanding the signal photons into two paths in displaced Sagnac interferometers, with the extra degree of freedom associated with the path of the photon corresponding to a qubit in each interferometer. The beamsplitters used in the interferometers are hybrids, with half of their surface a PBS and the other half a 50:50 beamsplitter (BS). The signal photons are input through the PBSs, so that their paths are correlated with their polarizations and the graph state is extended by a qubit at each end, creating a five-qubit linear graph. This is equivalent to the resource state shown in Figure~\ref{expsetup}~b up to local complementation operations~\cite{Bell13}, which are carried out using additional waveplates and a relabelling of the interferometer paths to the Pauli X basis (see Methods). The five-qubit graph state generated in the experiment and shown in Figure~\ref{expsetup}~b is given explicitly by
\bqa
&&\hskip-0.7cm \ket{\Psi}= \frac{1}{2\sqrt{2}} \big[\ket{-_y}_{0}\big(\ket{++0}\ket{-_y}-i\ket{++1}\ket{+_y}+i\ket{--0}\ket{-_y} +\ket{--1}\ket{+_y}\big)_{1234} \nonumber \\
&&\hskip0.1cm +\ket{+_y}_{0}\big(\ket{++0}\ket{+_y}+i\ket{++1}\ket{-_y}-i\ket{--0}\ket{+_y} +\ket{--1}\ket{-_y}\big)_{1234} \big], \label{EQN: 5 party graph state}
\eqa
where the eigenstates of the Pauli $Y$ operator are $\ket{\pm_y}=\frac{1}{\sqrt{2}}\left(\ket{0}\pm i\ket{1}\right)$. To measure the path qubit in the Pauli $X$ basis, one path or the other is blocked inside the interferometer. To measure in the $Y$ or $Z$ basis, the paths are allowed to recombine at the BS surface with different relative phases. The polarization qubits are then measured using quarter-wave plate (QWP) - HWP - PBS chains, followed by silicon avalanche photodiode detectors (APDs), which enable any Pauli basis measurement to be performed~\cite{James01}.

\qquad We first checked the entanglement of the graph state using an entanglement witness~\cite{Toth2005}. In this case, it is possible to detect genuine multipartite entanglement (GME) in a linear cluster state using the correlations from only two local measurement bases. Since the five-qubit graph state is locally equivalent to a five-qubit linear cluster state, by making corresponding changes to the reference frames of the measurements we obtain the relevant witness (see Methods). The measurements are $X_0 X_1 X_2 X_3 X_4$ and $Y_0 Z_1 Y_2 Y_3 Z_4$, which lead to a witness value of $\langle {\cal W} \rangle=-0.15 \pm 0.03$. The error is calculated using a Monte Carlo method with Poissonian noise on the count statistics~\cite{James01}. The negative expectation value of the witness reveals the presence of GME and confirms that all qubits are involved in the generation of the graph state. We also obtain the fidelity of the experimental graph state with respect to the ideal case using seventeen measurement bases (see Methods) and find a fidelity of $F=0.70 \pm 0.01$.

\qquad Having characterised the resource state, we move on to testing its performance in carrying out secret sharing protocols. We consider qubit $0$ to belong to the dealer, and qubits $1,2,3$ and $4$ to players $1,2,3$ and $4$ respectively. It can be seen from Eq.~(\ref{EQN: 5 party graph state}) that the graph state is a maximally entangled state between the dealer and the players. Thus, its use for secret sharing can be thought of as analogous to the way a maximally entangled state is used for two player communication. When using it to share a classical secret, a random key can be established between the dealer and authorised sets of players, similar to entanglement-based quantum key distribution (QKD)~\cite{Marin13}. On the other hand, when using it to share a quantum secret it can be thought of as the entangled resource for teleporting a secret state from the dealer to the players. In both cases the shape of the graph state imposes restrictions on which sets of players can access the secret, giving the overall access structure for the secret sharing.

\vskip0.4cm
{\bf Classical secret sharing (CQ)}
\vskip0.2cm

In the CQ protocol the graph state in Eq.~(\ref{EQN: 5 party graph state}) is used to establish a random key which can be known only by the dealer and an authorised set of players \cite{HBB99,MS08}. In this sense it is similar to a secret key generation protocol: once the key is established it can be used to securely communicate between the dealer and the authorised set of players, even in the presence of eavesdroppers (making it an improvement on the Shamir-Blakely schemes~\cite{Sha79,Blakely79}, which require trusted channels). We will see that it also can be used as a subprotocol for secure QQ. As in entanglement based QKD, the players both measure in randomly chosen complementary bases, the correlations are then checked, and if sufficiently high the key can be trusted.

\qquad The dealer starts by measuring their qubit either in the Pauli $Y$ or $Z$ basis, chosen at random. Here, the state in Eq.~(\ref{EQN: 5 party graph state}), which is written in the Y basis for the dealer, can also be written in an alternative form with the dealer's qubit in the Z basis as
\bqa
\ket{\Psi}&=&\frac{1}{2\sqrt{2}}\big[\ket{0}_{0}\big(\ket{++0}\ket{0}+\ket{++1}\ket{1}+\ket{--0}\ket{1} +\ket{--1}\ket{0}\big)_{1234} \nonumber \\
&& +\ket{1}_{0}\big(\ket{++0}\ket{1}+\ket{++1}\ket{0}-\ket{--0}\ket{0} +\ket{--1}\ket{1}\big)_{1234}\big].  \label{EQN: 5 party graph state 2}
\eqa
Thus, the dealer's measurement projects the players' state into one of four states $\rho_{1234}^{i,j}$, where $j=(Z,Y)$ represents the dealer's basis choice and $i=(0,1)$ the dealer's measurement result, which can easily be calculated from Eqs.~(\ref{EQN: 5 party graph state}) and (\ref{EQN: 5 party graph state 2}). The dealer's result is used as the secret key and the task of the players is to make measurements to discriminate the states $\rho_{1234}^{i,j}$ and find $i$. They cannot do this perfectly without knowledge of the basis choice $j$, so they make choices based on a guess $j'$ for the basis used by the dealer. As in standard QKD, after the players measure their state, the basis choice $j$ is announced by the dealer. If the players' measurements were chosen differently, {\it i.e.} $j'\neq j$, the results are discarded and a sifted key is built up using the cases where the bases of the dealer and the players coincided. For a given basis choice $j$, a set of players is `unauthorised' if there is no measurement they can make to find $i$, and a set of players is `authorised' if they are able to perfectly find $i$ using a particular choice of measurement. Further details of the protocol and proof of security can be found in refs.~\cite{MS08,Marin13}.

\begin{figure}[t]
\centering
\includegraphics[width=13cm]{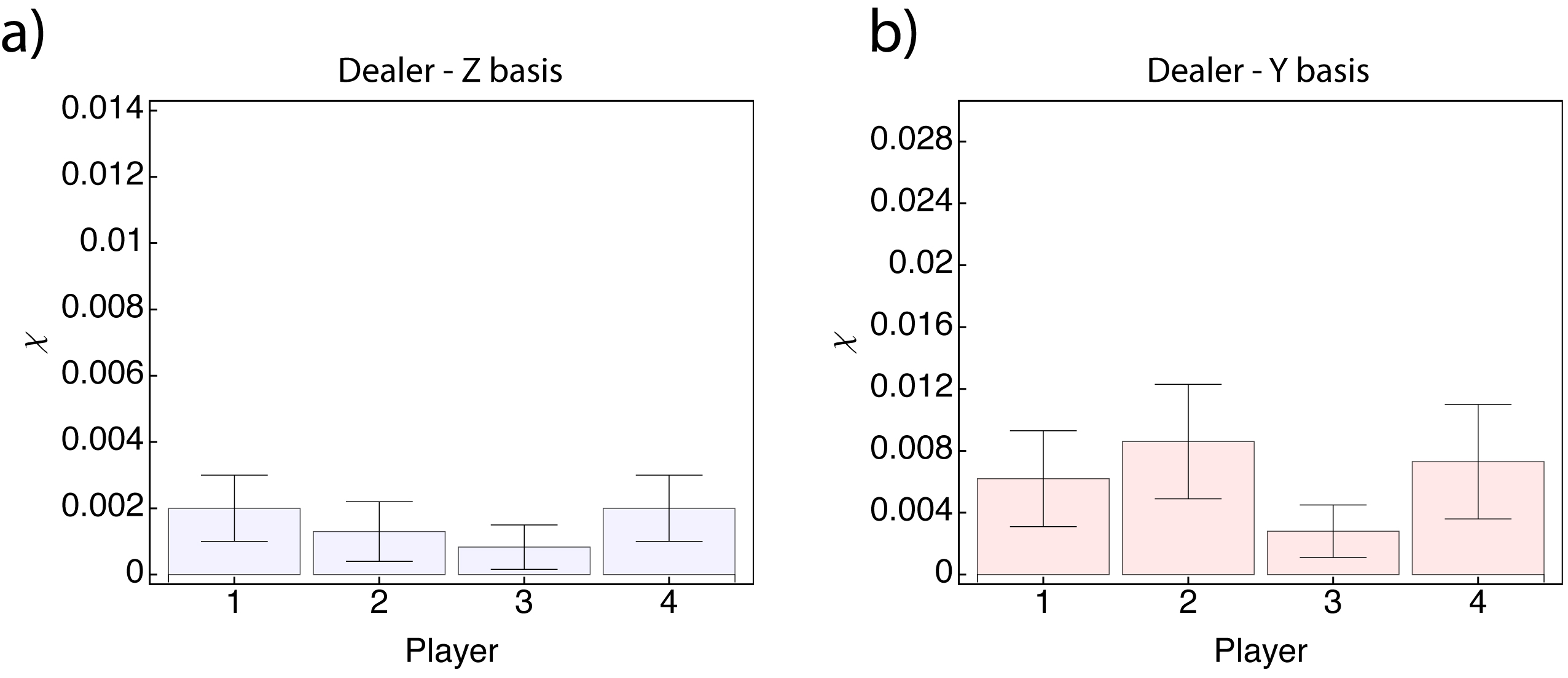}
\vskip1cm
\caption{}
\label{FigCQ1}
\end{figure}

\qquad To check whether a set of players $B$ using the five-qubit graph state can access the secret it is necessary to look at their reduced states $\rho_{B}^{i,j}$ given the dealer's result $i$ and basis choice $j$. For a particular basis $j$, the dealer is essentially encoding the classical information about their result $i$ into the state $\rho_{B}^{i,j}$, chosen with probability $p_{i,j}$ - the probability the dealer obtains result $i$ when measuring basis $j$. We denote this encoding of classical information as $\epsilon_j$. To quantify how well a set of players can access the dealer's results we then use the accessible information~\cite{Marin13}
\begin{equation}
\chi(\epsilon_j)=S(\rho_B^{j})-\sum_i p_{i,j}S(\rho_B^{i,j})
\end{equation}
where $S(\rho)$ is the von Neuman entropy of state $\rho$ and $\rho_B^{j}=p_{0,j}\rho_B^{0,j}+p_{1,j}\rho_B^{1,j}$. This quantifies the maximum possible information that the players can obtain about the dealer's results for a given basis choice $j$.
When $\chi(\epsilon_j)$ is zero there is no information that the players can obtain about the dealer's result, no matter which measurements they make.
It can be shown (see Methods) that using the state in Eq.~(\ref{EQN: 5 party graph state}) a single player cannot obtain any information about the dealer's result. That is, their reduced density matrix is independent of the dealer's result $i$, for both bases $j$. In Figure~\ref{FigCQ1} we have measured the reduced density matrices for each player (for each of the dealer's results) from the graph state generated in our experiment to obtain the accessible information $\chi$. One can see from Figure~\ref{FigCQ1}~a and \ref{FigCQ1}~b that the accessible information about the dealer's results are very close to zero for both the $Z$ and $Y$ bases, confirming that individual players have almost zero information about the dealer's results. When the dealer's result is not taken into consideration the fidelities of the single player reduced density matrices with respect to a maximally mixed state $I/2$ are $F=0.961 \pm 0.005$, $0.996 \pm 0.002$, $0.998 \pm 0.001$ and $0.996 \pm 0.002$ for players 1, 2, 3 and 4 respectively. When the dealer's result is taken into consideration, the 4 possible states for each player remain close to the maximally mixed state and lead to the low values of accessible information shown in Figure~\ref{FigCQ1}.

\begin{figure}[t]
\centering
\includegraphics[width=11cm]{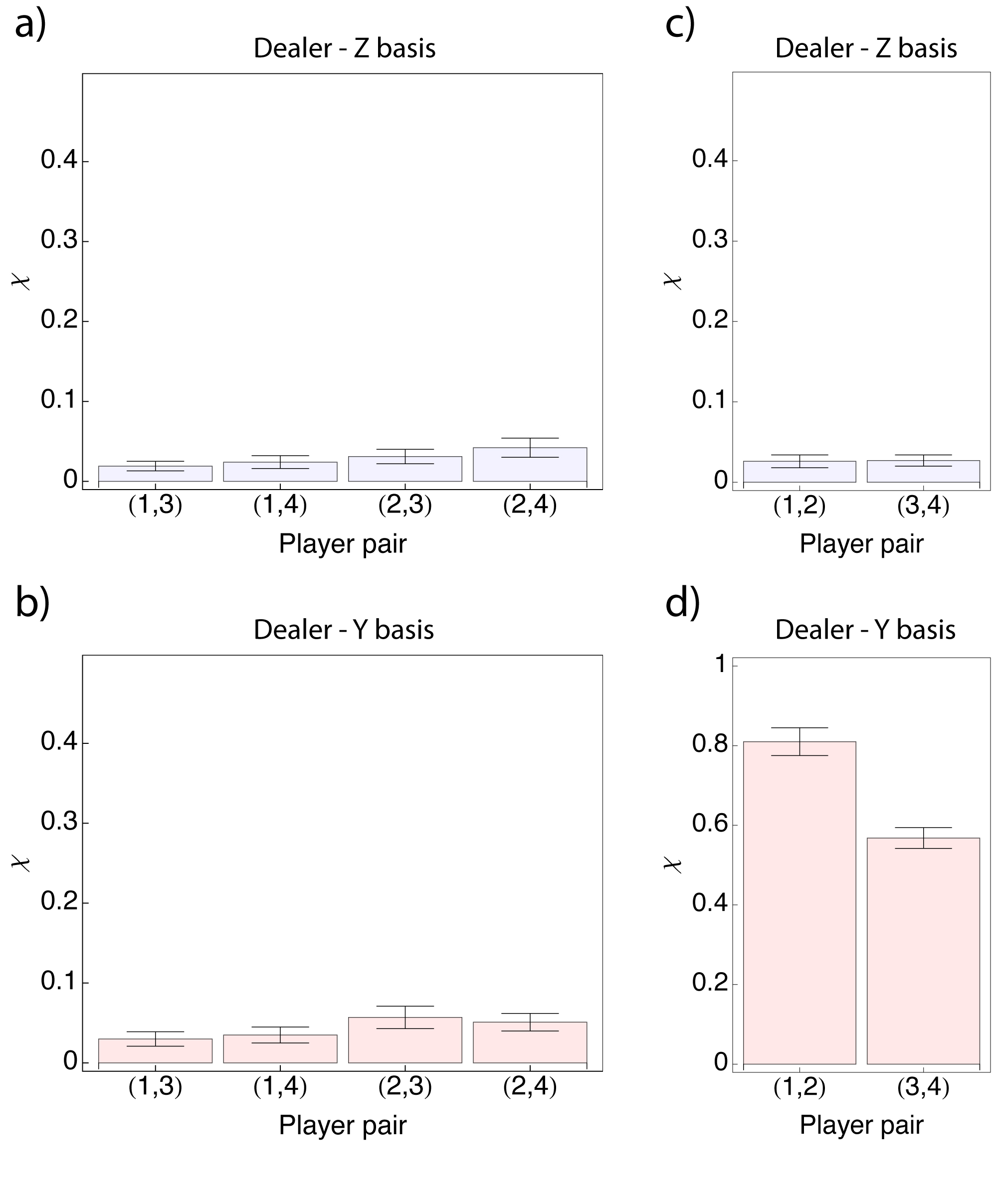}
\vskip1cm
\caption{}
\label{FigCQ2}
\end{figure}

\qquad For pairs of players, there are two cases, with the amount of information accessible different if the pair are adjacent: $(1,3)$, $(1,4)$, $(2,3)$ and $(2,4)$, or diagonally opposite: $(1,2)$ and $(3,4)$. It can easily be shown that a given pair of adjacent players cannot obtain any information about the dealer's result from the state in Eq.~(\ref{EQN: 5 party graph state}) (see Methods). One can see from Figure~\ref{FigCQ2}~a and \ref{FigCQ2}~b that the accessible information about the dealer's results are very close to zero for both the $Z$ and $Y$ bases using the graph state generated in our experiment, confirming that adjacent pairs of players have almost zero information about the dealer's results. On the other hand, when a pair of players are at opposite corners, it can be shown using Eqs.~(\ref{EQN: 5 party graph state}) and (\ref{EQN: 5 party graph state 2}) that when the dealer measures in the $Z$ basis they obtain no information, but when they measure in the $Y$ basis they obtain full information. For example, pair $(3,4)$ could do this by measuring in the bases $Z_3$ and $Y_4$. If the results are correlated, the dealer's result would be $1$, if they are anti-correlated it would be $0$. Similar conclusions can be found for the pair $(1,2)$. One can see from Figure~\ref{FigCQ2}~c and \ref{FigCQ2}~d that the measured accessible information about the dealer's results are very close to zero for the $Z$ basis, but not the $Y$ basis, confirming that opposite pairs of players can access information about the dealer's results.

\begin{figure}[t]
\centering
\includegraphics[width=13cm]{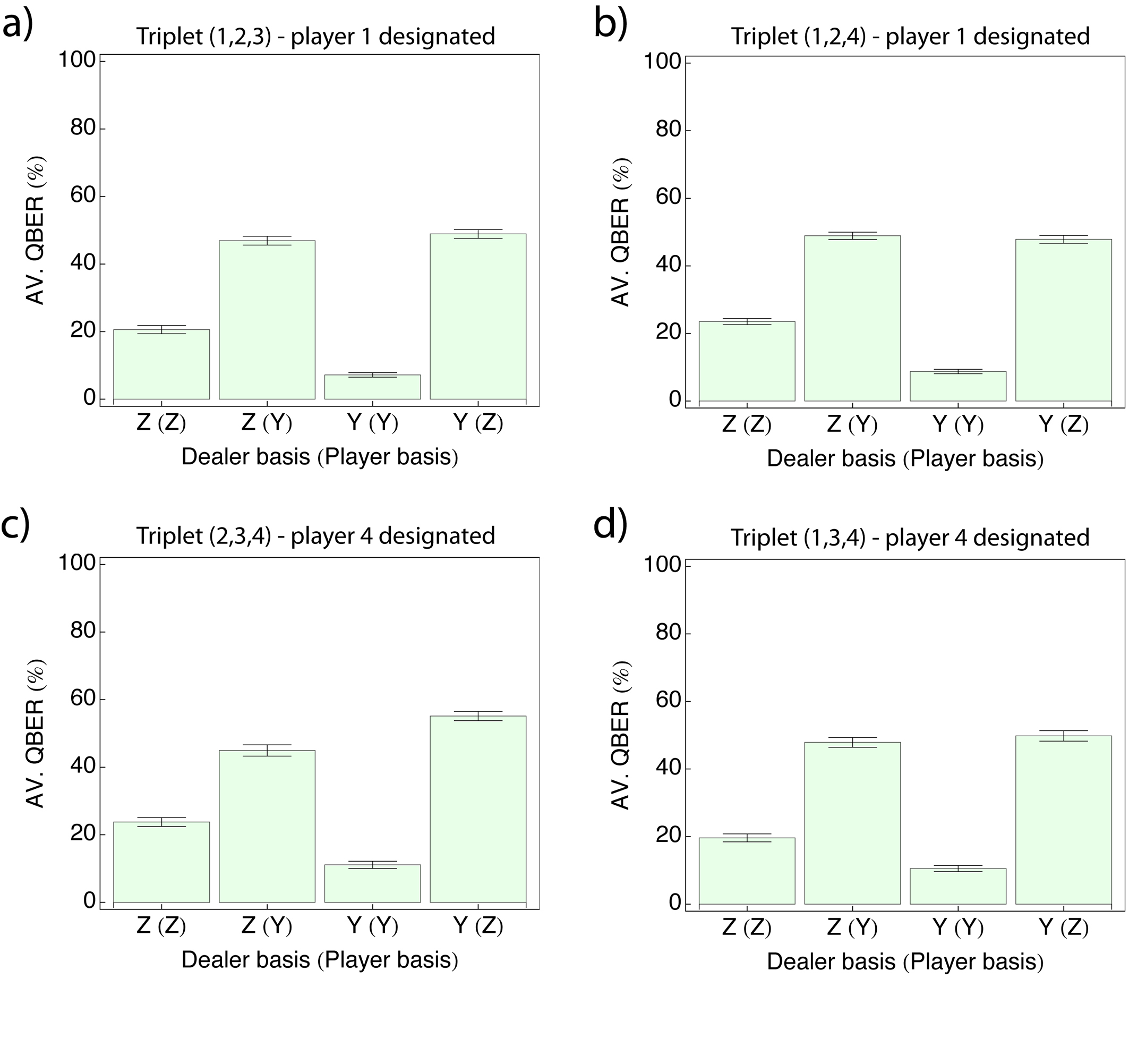}
\vskip1cm
\caption{}
\label{FigCQ3}
\end{figure}

\qquad Finally, for any three players it can be shown that in the ideal case they can access the dealer's measurement result perfectly for both the $Z$ and $Y$ bases. For example, if the dealer measures in the $Y$ basis, the triplet of players $(1,2,4)$ can obtain the result by measuring in the bases $Z_2$ and $X_4$ with the result of the dealer's outcome obtained from the measurement of player 1 in the $Y$ basis (after feedforward operations $X^{s_2}(XZ)^{s_4}Z$ are applied, where $s_i=(0,1)$ is the outcome of the measurement of qubit $i$). This also holds for the $Z$ basis of the dealer. These results can easily be checked by inspection of Eqs.~(\ref{EQN: 5 party graph state}) and (\ref{EQN: 5 party graph state 2}). The same retrieval process holds for any triplet of players by symmetry. The correlations within the graph state can therefore be used to establish a shared random key between the dealer and any set of three players. In Figure~\ref{FigCQ3} we show the measured quantum bit error rates (QBERs) for generating a shared random key for the four possible triplets of players: $(1,2,3)$, $(1,2,4)$, $(2,3,4)$ and $(1,3,4)$. The result of the dealer is obtained from the measurement of a designated player's qubit. Once the measurement basis is chosen, a non-zero QBER $p$ represents the action of the superoperator $\mathcal{E}^{(j)}(\rho^{i,j}_B) = p \rho^{i\oplus 1,j}_B + (1-p)\rho^{i,j}_B$. It is not difficult to see that for $p=50\%$ the accessible information will be zero, irrespective of $\rho^{i,j}_B$. One can see that the QBERs are low when the dealer and designated player measure in the same basis and close to 50\% when they use a different basis - corresponding to completely uncorrelated results. Taking the QBERs from both the $Z$ and $Y$ bases, when the dealer and designated player measure in the same bases, we almost reach the 11\% bound needed to establish a secure random key~\cite{Gisin2002}. We obtain $14 \pm 2 \%$, $16 \pm 2\%$, $18 \pm 2\%$ and $15 \pm 2\%$ for the triplets (1,2,3), (1,2,4), (2,3,4) and (1,3,4) respectively. Although our QBERs are just above the secure bound, the results demonstrate a first proof-of-principle for secure QKD with the access structure of the graph state. Note that in our experiment the players do not necessarily correspond to separate photons, as we are making use of hyperentanglement~\cite{Kwiat5}. This means that qubits (or players) one and two are embodied by one photon, and qubits three and four by another. For a pair of players that share a single photon, one can split up the access sets into their original form by allowing one player to control the measurement setting and readout of the path qubit, and the other to control the setting and readout of the polarization qubit.

\qquad In summary, using the graph state generated in our setup to share classical information via quantum channels (CQ) we have demonstrated a secret sharing scheme where a secret is distributed across four players such that any three can access the secret and any single player obtains no information. This is known as a \emph{ramp scheme} with parameters
$(3,1,4)$. Here, a ramp scheme $(k,k',n)$ enables the parameterising of any secret sharing scheme over $n$ players such that any set of $k$ or more players have perfect access to the secret and any set of $k'$ or fewer players have no access to the secret. If $k'=k-1$, then the scheme is called a $(k,n)$ threshold scheme.

\vskip0.4cm
{\bf Quantum secret sharing (QQ)}
\vskip0.2cm

We now show that our generated graph state can also be used to implement a (3,1,4) ramp scheme for sharing a {\it quantum} secret using the method described in ref.~\cite{MS08}. Furthermore, we also show how this ramp scheme can be upgraded to a (3,4) threshold scheme via hybrid quantum secret sharing (using both classical and quantum secret sharing)~\cite{Broadbent2009,Javelle11}. That is, any three players can access the quantum secret, but any fewer cannot. This is known to be impossible using a qubit pure quantum secret sharing protocol alone, {\it i.e.} without some classical mixing~\cite{MS08,Marin13}. Finally, we introduce and demonstrate a protocol which allows the sharing of a quantum secret over untrusted quantum channels between the dealer and the players.

\qquad In the QQ protocol a quantum state $\ket{\psi}=\alpha|0\rangle+\beta|1\rangle$ (the quantum secret) is encoded by the dealer onto the following four-qubit state shared by the players
\begin{equation} \label{EQN: QQ encoding}
\ket{\Phi}= \alpha |\phi\rangle_{1234} + \beta |\phi'  \rangle_{1234},
\end{equation}
where $|\phi \rangle_{1234}= 1/2(\ket{++00}+\ket{++11} + \ket{--01} + \ket{--10})_{1234}$ is a square graph state and $|\phi' \rangle_{1234}=Z_1Z_2Z_3Z_4|\phi \rangle_{1234}$.
The dealer achieves this encoding by teleporting in their secret state via a Bell measurement of the joint state of $\ket{\psi}$ and qubit 0 of the graph given in Eq.~(\ref{EQN: 5 party graph state})\cite{Marin13,Marin13b}. Alternatively, the dealer can directly prepare qubit $0$ of the graph in the secret state $\alpha|0\rangle_0+\beta|1\rangle_0$ and measure it in the $X$ basis with the feedforward operation $(Z_1Z_2X_3I_4)^{s_0}$ applied. In our experiment we implement this latter more compact approach for encoding the secret. The task of a set of players is then to access the secret quantum information.

\begin{figure}[t]
\centering
\includegraphics[width=15cm]{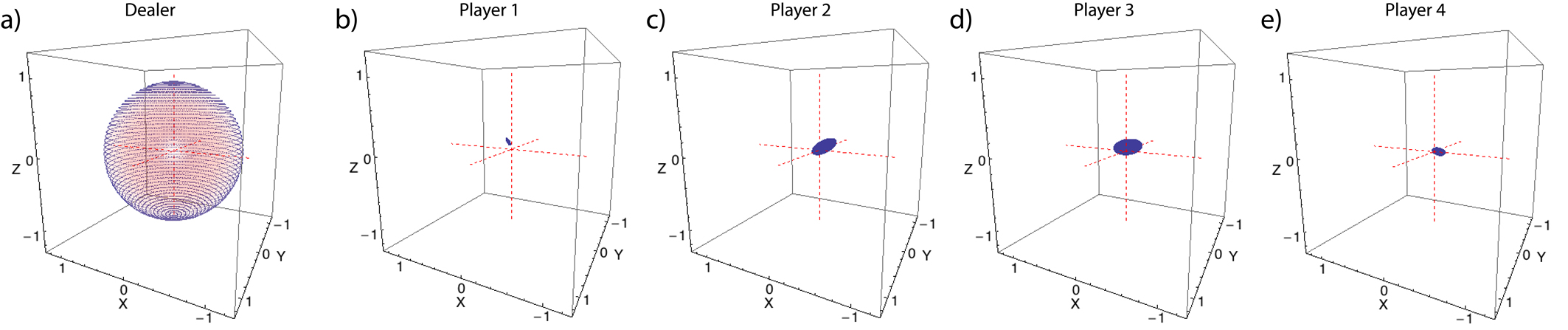}
\vskip1cm
\caption{}
\label{FigQQ1}
\end{figure}

\qquad In order to quantify the amount of information that can be accessed by a set of players $B$ we use the quantum mutual information of the reduced state shared by the dealer and the set of players, $\rho_{0,B}$ \cite{Marin13b}, which is given by
\begin{equation}
I(\rho_{0,B}) = S(\rho_{0}) + S(\rho_{B}) - S(\rho_{0,B}),
\end{equation}
where $\rho_0$ and $\rho_B$ are the reduced states of the dealer and players respectively. If $I(\rho_{0,B})$ is zero, the players obtain no information about the quantum secret.
For any single player it can be shown that the encoding in Eq.~(\ref{EQN: QQ encoding}) leads to a reduced density matrix that is maximally mixed, independent of the secret input qubit. In Figure~\ref{FigQQ1} we have used quantum process tomography and treated the mapping between the dealer and each player as a quantum channel for the secret qubit to be transferred over. Here, four probe states are used for the dealer's secret qubit, $\ket{0}$, $\ket{1}$, $\ket{+}$ and $\ket{+_y}$, which enable the reconstruction of the final Bloch sphere obtained by each of the players. One can see from Figure~\ref{FigQQ1} that using our generated graph state, all of the dealer's secret qubit states are transferred to states close to the maximally mixed state for the players. Furthermore, the measured mutual information between the dealer and each player, given in the caption, is consistently close to zero. Thus, a single player acting alone cannot obtain any information about the shared quantum secret.

\begin{figure}[t]
\centering
\includegraphics[width=15cm]{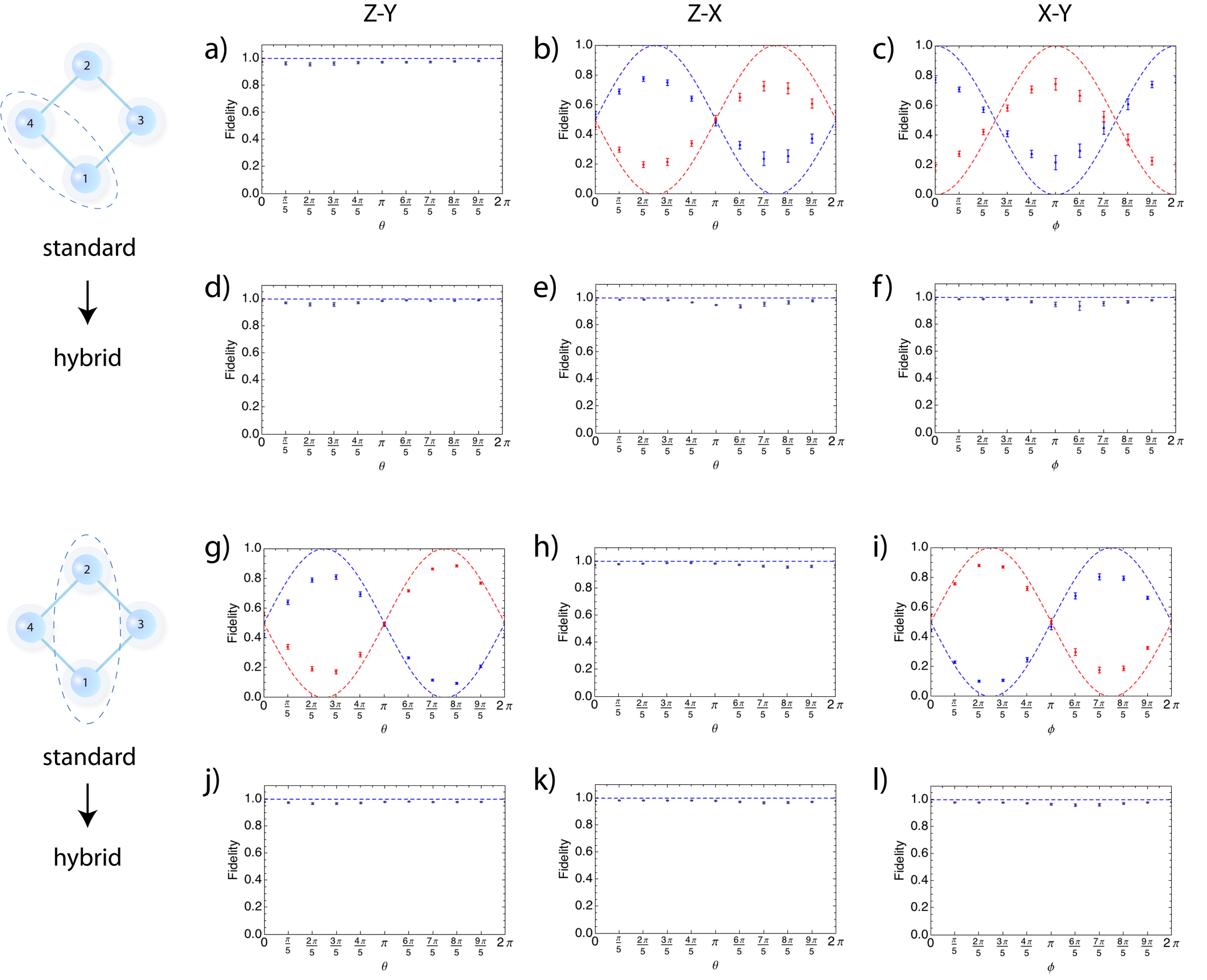}
\vskip1cm
\caption{}
\label{FigQQ2}
\end{figure}

\qquad For pairs of players the situation changes with regards to the amount of accessible information. When the players are adjacent, they obtain no information in the $Z$-$Y$ plane of the secret qubit's Bloch sphere, but can extract information in the $Z$-$X$ and $X$-$Y$ planes (see Methods). In Figure~\ref{FigQQ2}~a-c we show the experimental results from the player pair (1,4). Here, we plot the fidelity between the players' two-qubit state and fixed states as the dealer varies the angles in the respective planes of the Bloch sphere for their secret qubit. The fidelity gives an indication of how the state of the pair changes based on the dealer's input state. In Figure~\ref{FigQQ2}~a the fixed state is $I/4$ for the $Z$-$Y$ plane. In Figure~\ref{FigQQ2}~b and c, the fixed states are the orthogonal states $\frac{1}{4}(I+X\otimes X)$ and $\frac{1}{4}(I-X\otimes X)$ for both the $Z$-$X$ and $X$-$Y$ planes. The oscillations between the fixed orthogonal states show that some information about the dealer's qubit remains in the joint state of two players and depends on the plane the qubit is encoded into. We quantify the amount of information in this adjacent pair using the mutual information of the state shared by the dealer and the pair, measuring a value of $I=0.29 \pm 0.02$. This value is obtained from a three-qubit state tomography. On the other hand, when the players are opposite they obtain no information in the $Z$-$X$ plane, but can extract information in the $Z$-$Y$ and $X$-$Y$ planes. In Figure~\ref{FigQQ2}~g-i we show the experimental results from the player pair (1,2). In Figure~\ref{FigQQ2}~h the fixed state is $\frac{1}{4}(I+X\otimes X)$ for the $Z$-$X$ plane, while in Figure~\ref{FigQQ2}~g and i, the fixed states are the orthogonal states $\frac{1}{4}(I+X\otimes X+(Z \otimes Y+Y\otimes Z))$ and $\frac{1}{4}(I+X\otimes X-(Z \otimes Y+Y\otimes Z))/4$ for both the $Z$-$Y$ and $X$-$Y$ planes. Again, the oscillations between the fixed orthogonal states show that some information about the dealer's qubit remains in the joint state of two players. In this case, the mutual information of the state shared by the dealer and the pair is measured to be $I=0.62 \pm 0.02$, obtained from three-qubit state tomography.

\qquad To elevate the secret sharing QQ scenario to a threshold scheme, {\it i.e.} one where no two players can obtain any information, we use a hybrid protocol~\cite{Broadbent2009,Javelle11}. In this class of protocols, any $(k,k',n)$ ramp scheme can be elevated to a $(k,n)$ threshold scheme, and in fact all intermediate ramp schemes $(k,k'',n)$ for any $k'\leq k''\leq k-1$ can be achieved. In our case we can elevate the (3,1,4) ramp scheme to a (3,4) threshold scheme. The hybrid scheme uses, in addition to the QQ ramp scheme, a quantum one-time pad and classical secret sharing. That is, before the encoding the dealer applies a randomly chosen Pauli operation so that the state encoded is $X^xZ^z\ket{\psi}$, where $x,z$ are randomly chosen bits by the dealer. This state is then encoded and distributed, and the classical information $x,z$ is shared using classical secret sharing with ramp scheme parameters $(k,k'',n)$. Without the classical information the players will never be able to retrieve $\ket{\psi}$, but with the classical information, any $k$ can still access the information perfectly. In the present case, if the classical information is distributed using a classical $(2,4)$ secret sharing scheme, no two players can know its value. We check the performance of the hybrid protocol experimentally by applying randomly the operators $I$, $X$, $Z$ and $XZ$ to the dealer's qubit and measuring the resulting state of the pairs of players. In Figure~\ref{FigQQ2}~d-f we show the fidelity of the adjacent player's shared state with respect to the fixed state $I/4$ and in Figure~\ref{FigQQ2}~j-l we show the fidelity of the opposite player's shared state with respect to the fixed state $\frac{1}{4}(I+X\otimes X)$. One can see that when the dealer applies the hybrid encoding protocol, the information is almost completely removed from the state shared by pairs of players, as shown by the fidelities remaining constant over the angles of the planes.

\begin{figure}[t]
\centering
\includegraphics[width=14cm]{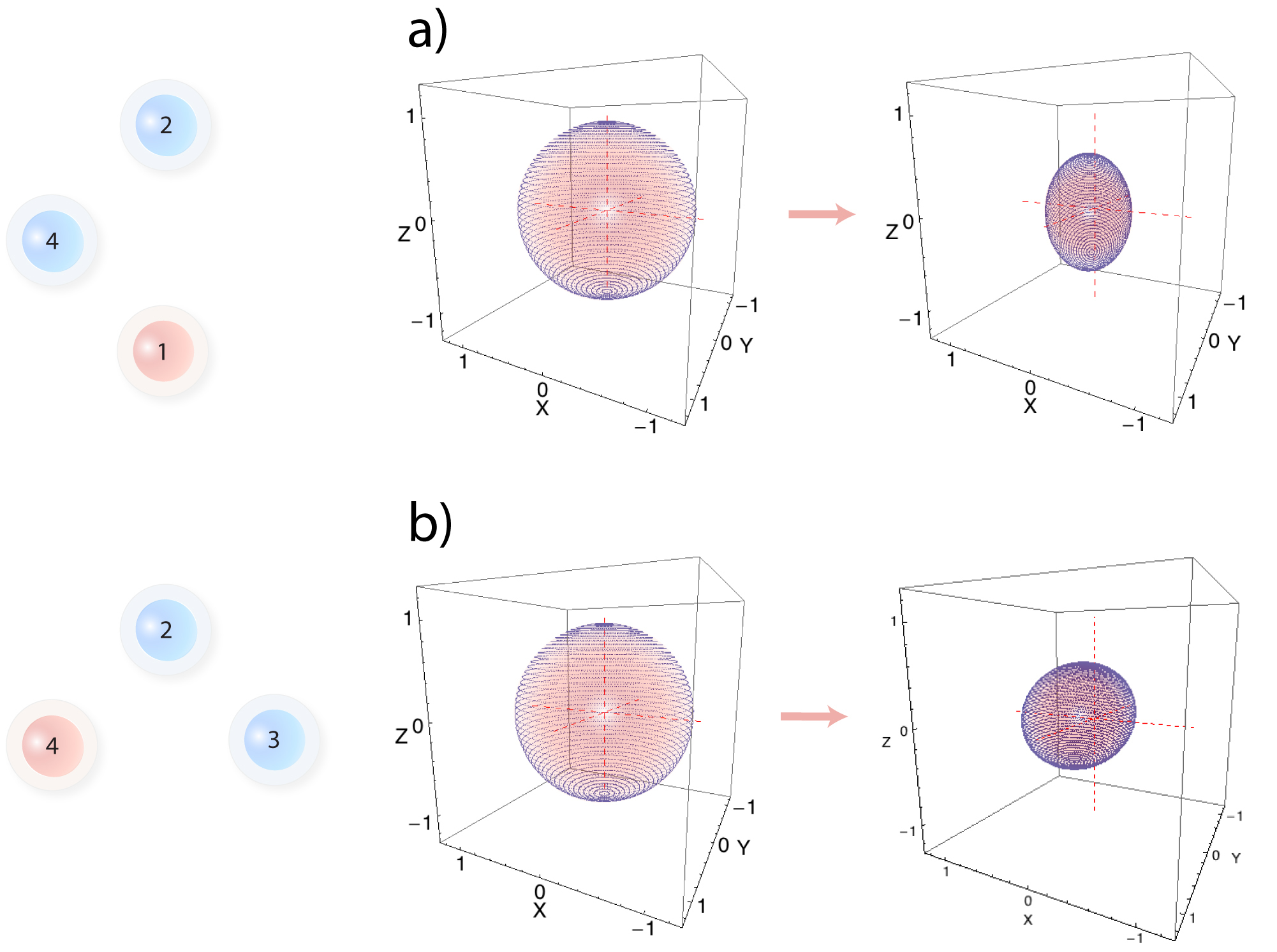}
\vskip1cm
\caption{}
\label{FigQQ3}
\end{figure}

\qquad For any set of three players the encoding in Eq.~(\ref{EQN: QQ encoding}) allows them to access the quantum secret perfectly. For example, if players $2$ and $4$ measure in $Z_2$ and $X_4$ respectively, the graph state is projected to one where the quantum secret resides on the qubit of player $1$ (up to a correction operation $X^{s_2}(ZX)^{s_4}Z$). The same is possible for any three players by symmetry. Thus, for the QQ protocol all three sets can access the secret. The same is true for the hybrid protocol, since the classical information of the one-time pad will be known by any set of three players and can easily be undone. In Figure~\ref{FigQQ3} we show the results from our generated graph state when the set of players (1,2,4) and (2,3,4) work to uncover the secret qubit shared by the dealer. Here, the designated player who retrieves the secret qubit is player 1 in Figure~\ref{FigQQ3}~a and player 4 in Figure~\ref{FigQQ3}~b. We again treat the mapping from the dealer to the designated player as a quantum channel and carry out quantum process tomography. One can see that in both sets of three players the secret quantum information is retrieved, although with some deformation of the Bloch sphere caused by the non-ideal graph state used in our experiment. However, the average fidelity for the shared qubit remains high with $\bar{F}_1=0.82 \pm 0.01$ and $\bar{F}_4=0.81 \pm 0.01$.


\newpage
\vskip0.4cm
{\bf Secure quantum secret sharing (SQQ)}
\vskip0.2cm

Finally, we introduce and demonstrate a new protocol for sharing a quantum secret over untrusted channels between the dealer and the players, which we denote by SQQ (for secure QQ). This is performed here for a $(3,4)$ threshold scheme, its extension to general access structures will be presented in ref.~\cite{Marin14}. The QQ protocol and the hybrid protocol work as long as the state used for encoding is the same as (or close to) that given in Eq.~(\ref{EQN: 5 party graph state}). However, if the channel from the dealer to the players is noisy or untrusted, this may not be the case. Thus, without knowing the initial secret that was sent, an authorised set of players cannot verify if they received it correctly or not. The SQQ protocol rectifies this problem by verifying that the state used is indeed that in Eq.~(\ref{EQN: 5 party graph state}), or close to it. Here, CQ measurements are used as a subprotocol to test the resource state (in a similar way to how a GHZ state can be tested using the verification protocol recently presented in ref.~\cite{Pappa12}).

\qquad The protocol works as follows: after generating and distributing the state, the dealer decides either to test it, or use it for quantum secret sharing, with probability $s$ and $1-s$, respectively. The dealer announces the choice about whether to test or use it publicly, and the dealer and players carry out their part of the test or the secret sharing scheme, respectively. The test is essentially an adapted version of the CQ protocol, which by checking the correlations verifies the state is the one desired. In the test, the dealer measures in either $X$, $Y$ or $Z$, or does not make any measurement, all with equal probability. They then announce their choice and the results publicly. A set of players $B$ which are checking the state, then do measurements depending on the dealer's measurement choice. The measurements used by the sets of three players are explicitly detailed below, along with a description of how the level of security is quantified.

\qquad It can be shown (see Methods) that if a given state $\rho$ shared between the dealer and players is used for secret sharing and the state $\omega$ that is retrieved by an authorised set of players has fidelity $f =\langle \psi | \omega | \psi \rangle$ with respect to the secret state $|\psi\rangle$, then the probability $P$ that the state $\rho$ passes the CQ test is related to the fidelity by
\begin{equation} \label{EQN: SQQ bound P}
f \geq 2P - 1.
\end{equation}
In other words, a state which passes the test with high probability will give a high fidelity when used for sharing the secret.

\qquad Furthermore, if we call $C_f$ the event that the protocol has not aborted and that the state $\rho$ was used for QQ, then we also show in the Methods that the probability $ P(C_f)$ of this event satisfies $f \geq \left(1- \frac{2s}{P(C_f)}\right)^{1/2}$. Thus, if the test is passed in the cases when the dealer announces they should test, then the players can be confident that when the dealer announces they should instead use the state, the secret quantum information retrieved will be of high fidelity.

\begin{figure}[t]
\centering
\includegraphics[width=13cm]{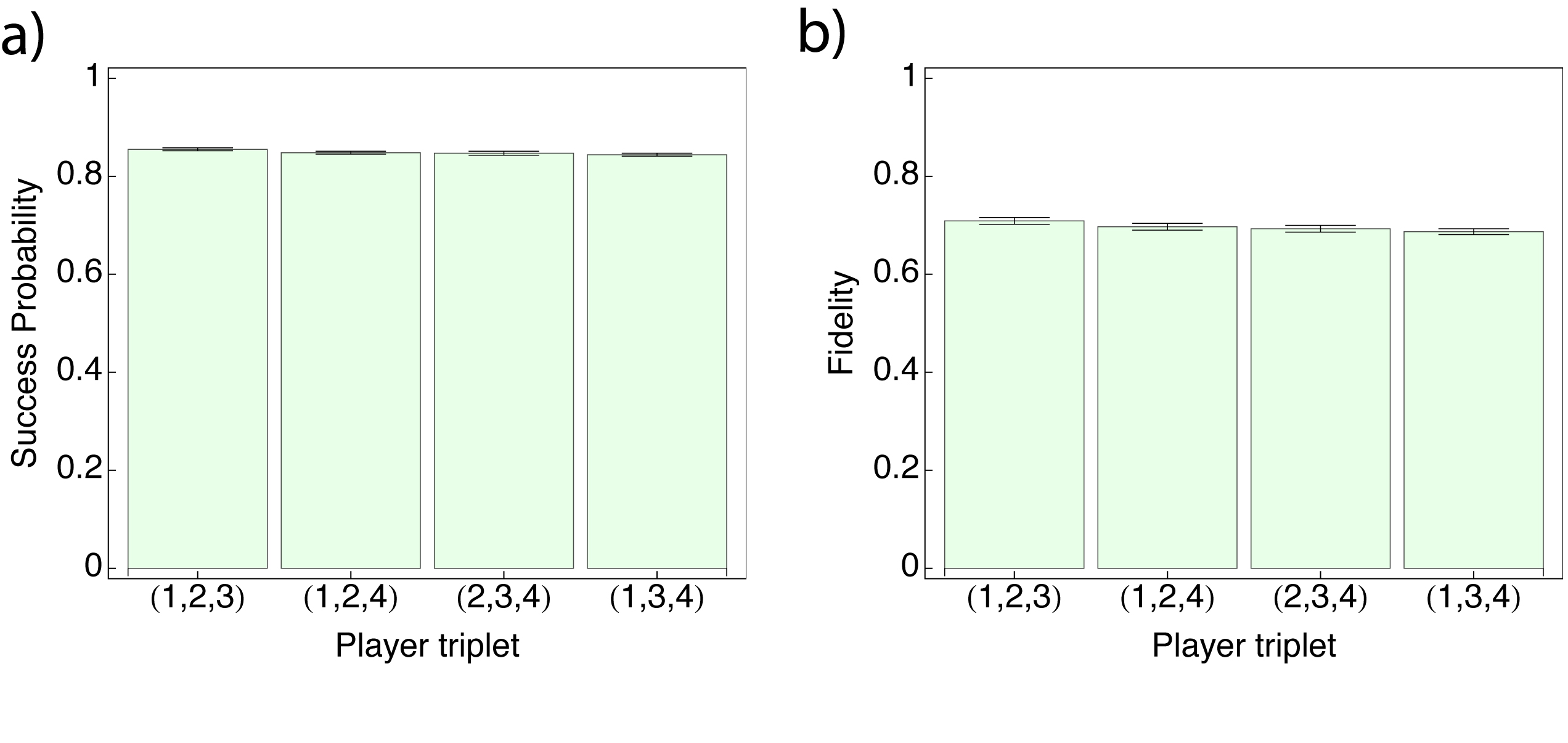}
\vskip1cm
\caption{}
\label{FigVQQ}
\end{figure}

\qquad As an example for our experimental graph state, we consider the set of players $(1,2,3)$, with the same holding for all sets by symmetry. The measurements for the test correspond to randomly measuring one of the following operators $Z_0Z_1Z_2X_3$, $Y_0Y_1Z_2I_3$, $Y_0Z_1Y_2I_3$, $X_0X_1I_2X_3$, $X_0I_1X_2X_3$, $I_0X_1X_2I_3$ or $Z_0Y_1Y_2X_3$ (see Methods). The test is passed if the measurement results for these operators are $+1$, $+1$, $+1$, $-1$, $-1$, $+1$ and $-1$, respectively. Based on the measured expectation values for these operator settings, in Figure~\ref{FigVQQ}~a we show the probability of our experimental state passing the test and in Figure~\ref{FigVQQ}~b we show the corresponding lower bounds on the fidelity, which are consistent with the fidelities measured previously in Figure~\ref{FigQQ3}. Thus, using the verified protocol we find that the probability of passing the test is fully consistent with the previously measured fidelity of the retrieved states for the three players.

\vskip0.5cm
\vskip0.8cm
{\Large {\bf Discussion}}
\vskip0.2cm

In this work we experimentally demonstrated the use of multipartite entangled graph states for classical (CQ) and quantum (QQ) secret sharing. We used a photonic setup to generate a five-qubit graph state and carried out the encoding, sharing and retrieval of classical and quantum secrets. In the CQ protocol we demonstrated the ability of the graph to share a classical random key, which can be used to securely share classical secrets, with an access structure of a $(3,1,4)$ ramp scheme. Here, the secret is shared between four players, such that any three can perfectly access the secret, yet no single player obtains any information at all. The QQ protocol achieves the same access structure for a quantum secret. However, with the integration of classical and quantum protocols, this access structure was then elevated to a $(3,4)$ threshold scheme for sharing a quantum secret, {\it i.e.} any three players can access the secret, but fewer have no information. This hybrid QQ protocol is a combination of classical secret sharing and the QQ protocol, which allows us to achieve an access structure known to be impossible with QQ alone. We also introduced and demonstrated a new protocol for sharing a quantum secret over untrusted channels, which we call SQQ. Taken together with the hybrid QQ protocol this highlights the power of integrating tasks using the graph state approach and enables us to achieve protocol parameters and security not possible with any single protocol. As more sophisticated ways of using graph states emerge, combining and demonstrating different sub-protocols in the way we have done here will become increasingly more relevant. The facility and flexibility of graph states for different quantum information processing tasks clearly propels them forward as a technology with great potential for future quantum networks.

\vskip0.5cm

{\Large {\bf Methods}}

{\bf Experimental setup}

The PCF sources used in the experiment are similar to those described in refs.~\cite{Halder09, Clark11}. When pumped by picosecond pulses from a Ti:Sapphire laser at $724$nm on the slow birefringent axis of the PCF, spontaneous four-wave mixing produces signal-idler photon pairs at $625$nm and $860$nm, polarized on the fast axis of the fibre. The cross-polarized phasematching scheme takes advantage of a turning point in the signal wavelength where it is locally independent of the pump wavelength, which has the effect of avoiding correlations between the signal and idler's spectra. This allows quantum interference to take place between photons from separate sources without the need for tight spectral filtering, which would reduce the collection efficiency.

\qquad To produce signal-idler pairs in a polarization Bell state, the PCF is set up in a Sagnac loop around a PBS and pumped in both directions. The axes of the fibre are twisted so that in the clockwise direction around the loop, the photon pairs polarized on the fast-axis emerge horizontally polarized, while for the counter-clockwise direction, photon-pairs emerge vertically polarized. When the two directions are recombined at the PBS, all the photon-pairs exit through the same output, so that the state of a single pair in this beam  is in a superposition $\frac{1}{\sqrt{2}}\left(\ket{HH}_{s1,i1}+e^{i\theta}\ket{VV}_{s1,i1}\right)$,
where the phase $\theta$ between the two directions can be tuned to zero using a birefringent compensator placed in the pump beam before the loop.

\qquad The other PCF source is pumped in a single direction so as to produce pairs without polarization entanglement. The idler is detected as a heralding photon while the signal photon is rotated to diagonal polarization $\frac{1}{\sqrt{2}}\left(\ket{H}_{s2}+\ket{V}_{s2}\right)$. This is then overlapped at the fusion PBS with the signal photon from the other source and we postselect events for the cases where one signal emerges from each PBS output. This implies that the two signal photons have the same polarization, or are in an even parity state, so that they have either both been transmitted or both been reflected at the PBS. The conditioned state is a three photon GHZ state $\frac{1}{\sqrt{2}}\left(\ket{HHH}_{s1,i1,s2}+\ket{VVV}_{s1,i1,s2}\right)$, which is converted to a linear graph state $\frac{1}{\sqrt{2}}\left(\ket{+0+}+\ket{-1-}\right)$ by waveplate rotations applied to the signal modes.

\qquad Each signal photon is then launched into a displaced Sagnac interferometer. Here, they are split at the PBS surface of the hybrid beamsplitters, and we label the transmitted paths the $\ket{0}$ states of the path qubits, and the reflected paths the $\ket{1}$ states. This results in the five-qubit state:
\begin{equation}
\begin{array}{l}
\frac{1}{2\sqrt{2}}\left(\ket{00}_{\rm pol, path}+\ket{11}_{\rm pol, path}\right)_{s1}\ket{0}_{i1}\left(\ket{00}_{\rm path, pol}+\ket{11}_{\rm path, pol}\right)_{s2}\\
\qquad +\frac{1}{2\sqrt{2}}\left(\ket{00}_{\rm pol, path}-\ket{11}_{\rm pol, path}\right)_{s1}\ket{1}_{i1}\left(\ket{00}_{\rm path, pol}-\ket{11}_{\rm path, pol}\right)_{s2},
\end{array}
\end{equation}
which is locally equivalent to a linear graph state and the target resource state, which can be written as:
\begin{equation}
\begin{array}{l}
\frac{1}{2\sqrt{2}}\left(\ket{++}_{12}+i\ket{--}_{12}\right)\ket{-y}_{0}\left(\ket{++}_{34}+i\ket{--}_{34}\right)\\
\qquad +\frac{1}{2\sqrt{2}}\left(\ket{++}_{12}-i\ket{--}_{12}\right)\ket{+y}_{0}\left(\ket{++}_{34}-i\ket{--}_{34}\right),
\end{array}
\end{equation}
where the eigenstates of the Pauli $Y$ operator are $\ket{\pm y}=\frac{1}{\sqrt{2}}\left(\ket{0}\pm i\ket{1}\right)$. The required local rotations are implemented by relabelling the transmitted and reflected interferometer paths to $\ket{+}$ and $\ket{-}$, applying HWP rotations to the signal polarizations, and a QWP rotation to the idler. Tilted glass plates in each path are used for the relative phase-shifts in the interferometers.

\qquad In order to experimentally implement the removal or tracing out of a path qubit (corresponding player does not take part in the secret sharing), the glass plate was removed from one path, so that the two path-lengths would differ by more than a coherence length. Hence the paths are incoherently recombined at the BS surface before going to polarization analysis. This allowed the photon's polarization to still be detected, but no information was gained about the path. On the other hand, to remove a polarization qubit, the PBS was taken away from the polarization analysis, so that the path information was still detected, but no polarization information was measured.

\qquad The two-fold coincidence rates collected from individual sources were around 9000 per second. Four-fold coincidences where the fusion succeeded, between the three entangled photons and the one herald photon, were $\sim0.25$ per second. Generating entanglement relies upon the signal photons from separate sources being indistingishable when they are overlapped at the PBS, otherwise the fusion can only leave an incoherent mix of possibilities~\cite{Bell12}. When the relative arrival time of the signal photons was varied, with the measurement bases set appropriately, an anti-dip was seen at zero-delay with visibility $\sim62\%$. This indicates there are some distinguishability issues which will degrade the quality of the state, which mainly result from inhomogenity along the length of the PCF sources.

\vskip0.2cm
{\bf Resource characterisation}

For the five-qubit graph state we use the following entanglement witness on qubits 0, 1, 2, 3, and 4
\bqa
{\cal W}&=&\frac{9}{4}I-\frac{1}{8}(\tilde{X}\tilde{X}II\tilde{X}+\tilde{X}\tilde{X}I\tilde{X}I+I\tilde{X}\tilde{X}\tilde{X}\tilde{X}+I\tilde{X}\tilde{X}II+\tilde{X}I\tilde{X}I\tilde{X}+\tilde{X}I\tilde{X}\tilde{X}I \nonumber \\
&&+III\tilde{X}\tilde{X})-\frac{1}{4}(IZ\tilde{Y}\tilde{Y}Z+\tilde{Y}Z\tilde{Y}II+\tilde{Y}II\tilde{Y}Z),
\eqa
where $\tilde{O}$ corresponds to measurements in the $O$ basis with the eigenstates swapped. This is a locally rotated version of the witness given in ref.~\cite{Toth2005} for a five-qubit linear cluster state and takes into account the required local complementation operations~\cite{Bell13}.

\qquad To obtain the fidelity for the five-qubit graph state we decompose the fidelity operator into a summation of products of Pauli matrices as
\bqa
F&=&\ketbra{\Psi}{\Psi}=\frac{1}{32}(1+IXXII-XXIXI-XIXXI-XXIIX-XIXIX \nonumber \\
&& +IIIXX+IXXXX+XYYYY+YZYII+YZYXX+YYZII \nonumber \\
&& +YYZXX-XZZYY-ZYYXI-ZYYIX-ZXIYY-ZIXYY \nonumber \\
&&+ZZZXI+ZZZIX+YIIZY+YXXZY+IZYZY+IYZZY \nonumber \\
&&+YIIYZ+YXXYZ+IZYYZ+IYZYZ-XYYZZ+XZZZZ \nonumber \\
&&+ZXIZZ+ZIXZZ).
\eqa
Calculating the expectation value of this operator requires 17 unique measurement bases: $XXXXX$, $YXXYZ$, $YXXZY$, $ZXXYY$, $ZXXZZ$, $XYYYY$, $XYYZZ$, $ZYYXX$, $YYZYZ$, $XYZZY$, $YYZXX$, $YZYYZ$, $ZZYZY$, $YZYXX$, $XZZYY$, $XZZZZ$ and $ZZZXX$.

\newpage
\vskip0.2cm
{\bf Classical secret sharing}

It can be seen by inspection of Eq.~(\ref{EQN: 5 party graph state}) that the reduced density matrix of any player $a$ is $\rho_a^{i,j}=I/2$ for all basis choices $j$ and results $i$ of the dealer's measurements. From this it can easily be checked that $\chi(\epsilon_j)=0$ for all bases $j$. Hence, no single party can obtain any information.

\qquad For a pair of players $(a,b)$ that are adjacent, one can easily check from Eq.~(\ref{EQN: 5 party graph state}) that $\rho_{a,b}^{i,j}=I/4$, $j=Z,Y$. Hence, no information can be extracted and $\chi(\epsilon_j)=0$ for all bases $j$. For a pair of players $(a,b)$ that are opposite it can easily be seen from Eq.~(\ref{EQN: 5 party graph state}) that for $j=Y$ they can access the result by measuring one qubit in $Y$ and the other in $Z$. Hence, $\chi(\epsilon_Y)=1$. It can also be shown that $\rho_{a,b}^{i,Z}=(1/2)(|++\rangle\langle++| +|--\rangle\langle--|)$ for both results $i$, so that no information can be extracted and $\chi(\epsilon_Z)=0$

\qquad Any triplet of players can access the secret, as discussed in the results section, which can be seen by inspection of Eqs.~(\ref{EQN: 5 party graph state}) and~(\ref{EQN: 5 party graph state 2}).

\vskip0.2cm
{\bf Quantum secret sharing}

Here we derive the access structure of the QQ protocol. The first step in the protocol is that the dealer generates and distributes the state in Eq.~(\ref{EQN: 5 party graph state}). We rewrite the state as follows
\begin{equation} \label{EQN: 5 party graph b}
\frac{1}{\sqrt{2}}(|0\rangle_0 |\phi\rangle_{1234} + |1\rangle_0 |\phi' \rangle_{1234}).
\end{equation}
This is used to teleport a secret state $|\psi\rangle = \alpha|0\rangle + \beta |1\rangle$ to the players. The dealer measures the secret qubit and their part of the state in Eq.~(\ref{EQN: 5 party graph b}) in the Bell basis and announces the results publicly. In the retrieval step the authorised sets then apply the appropriate correction and the decoding operations. To study the accessibility of the quantum information we ignore the correction step and assume it is always the good result where no correction is required - if a set of players cannot access the secret for the corrected state, then they cannot access it for the uncorrected state. Similarly, if they can, knowing the results of the dealer's measurement allows them to do the correction afterwards.
Thus, consider the secret teleported to the players, giving the state
\bqa
&&\hskip-0.5cm \alpha |\phi\rangle_{1234} + \beta |\phi' \rangle_{1234}=\big( |+\rangle_1 |+\rangle_2 |0\rangle_3 (\alpha|0\rangle_4 - \beta|1\rangle_4)+ |+\rangle_1 |+\rangle_2 |1\rangle_3 (\alpha|1\rangle_4 - \beta|0\rangle_4)\nonumber \\
&&\hskip1.5cm + |-\rangle_1 |-\rangle_2 |0\rangle_3 (\alpha|1\rangle_4 + \beta|0\rangle_4)+  |-\rangle_1 |-\rangle_2 |1\rangle_3 (\alpha|0\rangle_4 + \beta|1\rangle_4)\big)/4. \label{EQN: QQ encoding Expansion}
\eqa
Note that this state is cyclically symmetric amongst the four players, according to the symmetry of the graph, in this case a square. It can be seen from Eq.~(\ref{EQN: QQ encoding Expansion}) that any single player $a$ has the reduced density matrix $\rho_a=I/2$, thus they cannot access any information. This is quantified by considering the reduced state of Eq.~(\ref{EQN: 5 party graph b}) for $\rho_{0a}=I/4$, so that the mutual information $I(\rho_{0a})=0$.

\qquad For two adjacent players $a$ and $b$ we have from Eq.~(\ref{EQN: QQ encoding Expansion})
\begin{align}
\rho_{ab} = & \frac{1}{2}\big( \ket{0}_a\bra{0}\otimes \left( XZ\ket{\psi}_b\bra{\psi}ZX +Z\ket{\psi}_b\bra{\psi}Z \right) \nonumber \\
&+  \ket{1}_a\bra{1}\otimes \left( X\ket{\psi}_b\bra{\psi}X + \ket{\psi}_b\bra{\psi} \right) \big).
\end{align}
From this we find that they can obtain some information as follows: player $a$ measures in the $Z$ basis (the result of which we denote $s_a$), and then tells player $b$ the outcome. Player $b$ then performs the correction $Z^{1\oplus s_a}$ and we are left with the state on player $b$ as $\rho_b= \left( X\ket{\psi}_b\bra{\psi}X + \ket{\psi}_b\bra{\psi} \right) /2$. Thus in some cases the full information can be retrieved and in other cases only partially, depending on the secret shared. For example, if the secret state is $\ket{\pm}$ then full information can be retrieved.

\qquad On the other hand, after teleportation two opposite players $a$ and $b$ share the state
\be
\rho_{ab} = \ket{A}_{ab}\bra{A} + \ket{B}_{ab}\bra{B} + i \sin(\theta) \sin(\phi) \left(\ket{A}_{ab}\bra{B} - \ket{B}_{ab}\bra{A} \right),
\label{EQN: QQ pair opposite}
\ee
where $\ket{A}=(\ket{++}_{ab} + \ket{--}_{ab})/\sqrt{2}$, $\ket{B}=(\ket{++}_{ab} - \ket{--}_{ab})/\sqrt{2}$, and we take the standard Bloch sphere parameterisation of the input qubit $\alpha=\cos(\theta/2)$ and $\beta=e^{i\phi}\sin(\theta/2)$. The players can retrieve information as follows: player $a$ measures in the $Z$ basis, obtaining result $s_a$, and player $b$ performs the correction operation $Z_b^{s_a\oplus 1}$. The resulting state is $\rho_b= \left( ZX\ket{\psi}_b\bra{\psi}XZ + \ket{\psi}_b\bra{\psi} \right) /2$. Thus in some cases the full information can be retrieved and in other cases only partially, depending on the secret shared. For example, if the secret state is $\ket{\pm_y}$ then full information can be retrieved.

\qquad For three players, it can easily be seen from the decomposition of Eq.~(\ref{EQN: 5 party graph b}) that if player $2$ measures $Z_2$ and player $4$ measures $X_4$, the secret can be retrieved on the qubit of player $1$ up to feedforward operations $X^{s_2}(XZ)^{s_4}Z$. Similar results hold for all sets of three players by symmetry of the graph state.

\vskip0.2cm
{\bf Hybrid Quantum secret sharing}

After the random application of the operators $I$, $X$, $Z$ and $XZ$ based on the results of a one-time pad, as well as the QQ encoding teleportation stage, the state of the players is
\begin{equation} \label{EQN: hybrid QQ encoding}
 \frac{1}{\sqrt{2}}X_L^x Z_L^z \left(\alpha |\phi\rangle_{1234} + \beta |\phi' \rangle_{1234}\right),
\end{equation}
where $X_L=Z_1Z_2X_3I_4$ and $Z_L=Z_1Z_2Z_3Z_4$. From the arguments in the previous section, players who cannot access the quantum secret in the QQ case cannot access it in this case too. However, players also cannot access anything when they do not know the values $x$ and $z$. This can be checked by looking at the reduced density matrices mixed over the values of $x$ and $z$. Thus, any two players not knowing $x$ and $z$ obtain no information, but any three knowing $x$ and $z$ can access the secret state perfectly. Sharing the classical information of $x$ and $z$ via a $(3,4)$ Shamir-Blakely~\cite{Sha79,Blakely79} classical secret sharing scheme achieves this exactly. Note, we are assuming authenticated classical channels, as in all our schemes. However, to use the  Shamir-Blakely~\cite{Sha79,Blakely79} secret sharing scheme one also requires a trusted channel. If one does not trust the classical channels one could use a CQ scheme to send this information, or indeed the Shamir-Blakely scheme plus multiple standard two party QKD.

\vskip0.2cm
{\bf Secure quantum secret sharing details}

We now present the verified SQQ protocol and its proof in more detail. We exemplify the protocol for our state with accessing set $B=(1,2,3)$. The same steps can be performed by symmetry for all sets of three players.
\begin{enumerate}
\item \label{protocol Verif QQ step 1}The dealer distributes the players' qubits of the entangled graph state, {\it i.e.} the channel state in Eq.~(\ref{EQN: 5 party graph state}).
\item The dealer randomly decides that they will carry out: (a) the protocol $CQ_{test}^B$, or (b) the QQ protocol,  with probabilities $s$ and $1-s$ respectively, and announces the choice to all players.
\begin{enumerate} \item {\it $CQ_{test}^B$ defined for an authorised set $B=(1,2,3)$}.
 \begin{enumerate}
\item The dealer chooses randomly which of the seven measurements below should be performed for the test, announcing the choice and results of their part of the measurement (in the case where they are asked to measure $I_0$ they output the result $+1$).
\begin{align}
M_1 &=  Z_0 Z_1 Z_2 X_3 \nonumber\\
M_2 &=  Y_0 Y_1 Z_2 I_3 \nonumber\\
M_3 &=  Y_0 Z_1 Y_2 I_3 \nonumber\\
M_4 &= -X_0 X_1 I_2 X_3 \nonumber\\
M_5 &= -X_0 I_1 X_2 X_3 \nonumber\\
M_6 &=  I_0 X_1 X_2 I_3 \nonumber\\
M_7 &= -Z_0 Y_1 Y_2 X_3 \nonumber
\end{align}
The minus signs can be interpreted as meaning that the product of outputs should ideally be minus one.
\item To comply, players in $B$ perform their parts of the measurement chosen by the dealer. They then check their correlations by communicating amongst themselves. If the product of outcomes of the dealer and all $B$ is 1  (or $-1$ according to the sign of the measurement), they give the response ``pass'', otherwise ``fail''.
\item If ``pass'' is returned, proceed again to step 1 of the protocol, otherwise abort.
\end{enumerate}
\item Carry out standard QQ defined for the authorised set $B=(1,2,3)$
 \begin{enumerate}
\item The dealer measures the quantum secret and their part of the shared channel state in the Bell basis and announces the result.
\item Players $B$ perform the corrections and decoding operation.
\end{enumerate}
\end{enumerate}
\end{enumerate}

We now give a proof of the security for the QSS protocol, {\it i.e.} we prove the fidelity bounds with respect to passing the test. Carrying out the protocol $CQ_{test}^B$ as defined above for players $(1,2,3)$ is equivalent to performing a POVM $\{M_{pass},M_{fail}\}$, where $M_{pass}$ is the sum of all the $+1$ projections for the measurements performed in the $CQ_{test}^B$, which can easily be seen to give
\begin{eqnarray}
M_{pass} =\sum_i\frac{M_i + I}{2}=  \frac{I+\Gamma}{2},
\end{eqnarray}
where $\Gamma = \left(|g\rangle_{0123}\langle g| + I_0 Z_1 Z_2 Z_3|g\rangle_{0123}\langle g|I_0 Z_1 Z_2  Z_3\right)$ is the projection onto a space where the QQ protocol works perfectly, and $|g\rangle_{0123}$ is the graph state of the subgraph of qubits 0, 1, 2 and 3. The probability $P$ of passing the $CQ_{test}$, given a state $\rho$, is then given by
\begin{eqnarray}
P = \tr (\rho M_{pass}) = \frac{1 + \tr(\rho \Gamma)}{2}.
\end{eqnarray}

\qquad Consider $\rho$ is now used instead to share a quantum secret $|\psi\rangle$ via the QQ protocol. If we denote $f=\langle \psi |\omega|\psi\rangle$ the fidelty of the decoded state $\omega$, then it follows that $f \geq\tr(\rho \Gamma)$, since any state in the subspace $\Gamma$ perfectly transports the secret, so the final fidelity can only be higher than the overlap with this space, giving Eq.~(\ref{EQN: SQQ bound P}).

\qquad Following the logic in ref.~\cite{Pappa12}, if we denote $C_f$ the event that the certified protocol has not aborted and that the state $\rho$ was used for QQ such that it returns a decoded state with fidelity $f$ with the original secret, then it can be shown that the probability $P(C_f)$ of this event satisfies
\begin{equation}
P(C_f) \leq \frac{2s}{(1-f^2)}.
\end{equation}
This implies that if the test passes, then the fidelity of the output state is high. This relationship is demonstrated in the results section.  A generalisation of this protocol, with a more detailed and general proof can be found in ref.~\cite{Marin14}.

\vskip0.8cm
\begin{changemargin}{1cm}{0cm}


\end{changemargin}
\vskip1.8cm


{\scriptsize {\bf Acknowledgments} This work was supported by the UK's Engineering and Physical Sciences Research Council, ERC grant 247462 QUOWSS, EU FP7 grant 600838 QWAD, the National Research Foundation and Ministry of Education, Singapore, the Leverhulme Trust, the HIPERCOM (2011-CHRI-006) project and the Ville de Paris Emergences program, project CiQWii.}

{\scriptsize {\bf Competing interests statement} The authors declare that they have no competing financial interests.}

{\scriptsize {\bf Corresponding Authors} Correspondences and requests for materials should be addressed to \\ D. Markham ({\color{blue} \underline{damian.markham@gmail.com}}) or M. S. Tame ({\color{blue} \underline{markstame@gmail.com}}).}

{\scriptsize {\bf Author Contributions} B.A.B., D.M., D.A.H-M., A.M., J.G.R. and M.S.T. jointly conceived the graph state secret sharing scheme, the experimental layout and methodology. B.A.B. performed the experiments. M.S.T. led the project. All authors discussed the results and participated in the manuscript preparation.}

\newpage


{\myfontHel {\bf \myfontHel Figure 1. Experimental setup. a,} Setup used to generate the graph state resource for secret sharing. Two photonic crystal fibre sources are pumped using a Ti:Sapphire laser producing picosecond pulses at 724 nm. The first source produces a pair of photons in the state $\ket{H}_{i_1}\ket{H}_{s_1}$ and the second produces photons in the state $\frac{1}{\sqrt{2}}(\ket{H}_{i_2}\ket{H}_{s_2}+\ket{V}_{i_2}\ket{V}_{s_2})$. The signal photons from the first pair are rotated to the state $\ket{+}$ using a half-wave plate (HWP) and both signal photons are then fused using a polarizing beamsplitter. The polarizations of the signal photons are then rotated using HWPs to form the three-qubit linear cluster state $\frac{1}{\sqrt{2}}(\ket{+}_{s_1}\ket{H}_{i_2}\ket{+}_{s_2}+\ket{-}_{s_1}\ket{V}_{i_2}\ket{-}_{s_2})$, where the first idler photon is used as a trigger to verify a four-fold coincidence signifying the generation of the state. The path degree of freedom of the signal photons is then used to expand the resource to a five-qubit linear cluster state using a Sagnac interferometer, as shown in the dashed boxes and explained in the main text. Local complementation operations are then carried out in order to rotate the linear cluster into the graph state shown in panel {\bf \myfontHel b}, as detailed in ref.~\cite{Bell13}. {\bf \myfontHel b,} Diagram of the secret sharing scenario. Here, the vertices correspond to qubits initialized in the state $\ket{+}$ and edges correspond to controlled-phase gates, $C_Z={\rm diag}(1,1,1,-1)$, applied to the qubits.}

\vskip1cm

{\myfontHel {\bf \myfontHel Figure 2. Classical secret sharing using a quantum resource - single player. a,} Accessible information $\chi$ for single players when the dealer measures in $Z$. {\bf \myfontHel b,} Accessible information $\chi$ for single players when the dealer measures in $Y$. In both panels it is clear that when acting alone the players have almost zero information about the state the dealer has prepared. All error bars in the figures are calculated using a Monte Carlo method with Poissonian noise on the count statistics~\cite{James01}.}

\vskip1cm

{\myfontHel {\bf \myfontHel Figure 3. Classical secret sharing using a quantum resource - two players. a (c),} Accessible information $\chi$ for pairs of adjacent (opposite) players when the dealer measures in $Z$. {\bf \myfontHel b (d),} Accessible information $\chi$ for pairs of adjacent (opposite) players when the dealer measures in $Y$. From the panels it is clear that only opposite players have some information about the state the dealer has prepared when the dealer measures in the $Y$ basis.}

\vskip1cm

{\myfontHel {\bf \myfontHel Figure 4. Classical secret sharing using a quantum resource - three players. a-d,} Average quantum bit error rate (QBER) for the bit retrieved by the three player triplets when working together, with the bit retrieved by a designated player. For the designated player, the measured bit outcomes are correlated in the same bases as the dealer (ideally zero QBER) and uncorrelated in the opposite bases (ideally 50\% QBER).}

\vskip1cm

{\myfontHel {\bf \myfontHel Figure 5. Quantum secret sharing - one player.} Single-qubit Bloch spheres for individual players. Here, each Bloch sphere represents the output qubit states for an arbitrary state encoded by the dealer.  {\bf \myfontHel a,} Original Bloch sphere of states encoded by the dealer. {\bf \myfontHel b,} Player 1 Bloch sphere. {\bf \myfontHel c,} Player 2 Bloch sphere. {\bf \myfontHel d,} Player 3 Bloch sphere. {\bf \myfontHel e,} Player 4 Bloch sphere. One can see the Bloch spheres all correspond to an almost completely mixed state $I/2$ for the dealer's input states. The corresponding mutual information shared between the dealer and players 1, 2, 3 and 4 is $I=0.005 \pm 0.001$, $0.009 \pm 0.002$, $0.013 \pm 0.003$ and $0.009 \pm 0.003$. The process fidelity of the channel describing the mapping of the dealer's states to the players states is given by~\cite{Jozsa1994} $F_p=({\rm Tr}\sqrt{\sqrt{\chi_{\rm exp}}\chi_{\rm id}\sqrt{\chi_{\rm exp}}})^2$, where $\chi_{\rm id}$ is an ideal maximally mixed channel and $\chi_{\rm exp}$ is the experimentally reconstructed one. From this definition we obtain process fidelities of $0.989 \pm 0.002$, $0.973 \pm 0.043$, $0.980\pm 0.005$, and $0.975 \pm 0.056$ for players 1, 2, 3 and 4, respectively. Thus, in the one player case, the players have almost no information about the state the dealer has shared using the graph state.}

\vskip1cm

{\myfontHel {\bf \myfontHel Figure 6. Quantum secret sharing - two players.} Fidelities with respect to fixed reference states for the two-qubit states shared by two players as the dealer encodes qubit states into the graph along three orthogonal Bloch sphere planes $Z$-$Y$, $Z$-$X$ and $X$-$Y$ (parameterized by the canonical angles $\theta$ and $\phi$). {\bf \myfontHel a-f,} Fidelities for adjacent players (1 and 4), with panels {\bf \myfontHel a-c} showing the standard encoding scheme, where the mutual information between the dealer and the pair of players is $I=0.29 \pm 0.02$. Panels {\bf \myfontHel d-f} show the hybrid encoding scheme used to remove information in all three planes. {\bf \myfontHel g-l,} Fidelities for opposite players (1 and 2), with panels {\bf \myfontHel a-c} showing the standard encoding scheme, where the mutual information between the dealer and the pair of players is $I=0.62 \pm 0.02$ and panels {\bf \myfontHel d-f} showing the hybrid encoding scheme. The fixed reference states in the panels are as follows: In {\bf \myfontHel a, d, e} and {\bf \myfontHel f}, the fixed state is $I/4$. In {\bf \myfontHel b} and {\bf \myfontHel c}, the fixed states are the orthogonal states $\frac{1}{4}(I+X\otimes X)$ (red) and $\frac{1}{4}(I-X\otimes X)$ (blue). In {\bf \myfontHel h, j, k} and {\bf \myfontHel l}, the fixed state is $\frac{1}{4}(I+X\otimes X)$. In {\bf \myfontHel g} and {\bf \myfontHel i}, the fixed states are the orthogonal states $\frac{1}{4}(I+X\otimes X+(Z \otimes Y+Y\otimes Z))$ (red) and $\frac{1}{4}(I+X\otimes X-(Z \otimes Y+Y\otimes Z))/4$ (blue). The oscillations between the fixed orthogonal states show that some information about the dealer's qubit remains in the joint state of two players - depending on the plane the qubit is encoded into and quantified by the mutual information values $I$. However, when the dealer applies the hybrid encoding, the information is almost completely removed as shown by the fidelities remaining constant over the angles of the planes.}

\vskip1cm

{\myfontHel {\bf \myfontHel Figure 7. Quantum secret sharing - three players. a,} Bloch sphere of an arbitrary qubit shared by the dealer to players 1, 2 and 4, with the secret residing on the qubit of player 1. The average fidelity for a shared qubit is $\bar{F}=0.82 \pm 0.01$. {\bf \myfontHel b,} Bloch sphere for players 2, 3 and 4, with the secret residing on the qubit of player 4. The average fidelity for a shared qubit in this case is $\bar{F}=0.81 \pm 0.01$. In both, the spheres are slightly squashed due to the non-ideal graph state resource used in the experiment.}

\vskip1cm

{\myfontHel {\bf \myfontHel Figure 8. Verified quantum secret sharing protocol. a,} Expected success probabilities for the protocol being carried out between the dealer and different sets (triplets) of three players. The success of the verified protocol (see text for details) allows the dealer to verify the access structure of the graph resource and carry out verified quantum secret sharing. {\bf \myfontHel b,} Lower bound on the fidelity of the state shared by the players with respect to the ideal graph state.}

\newpage


\phantom{x}
\newpage

\phantom{x}
\newpage

\phantom{x}
\newpage

\phantom{x}
\newpage

\phantom{x}
\newpage

\phantom{x}
\newpage

\phantom{x}
\newpage

\phantom{x}

\end{document}